# Direct Observation of Large Flexoelectric Bending at the Nanoscale in Lanthanide Scandates

Pratik Koirala, Christopher Mizzi and Laurence D. Marks*

Department of Materials Science and Engineering, Northwestern University, Evanston, IL 60208, USA

**Abstract**

Large bending of materials can occur at the nanoscale in response to an electric polarization, what is called the flexoelectric effect, but to date this has not been observed directly. We report the direct observation of large flexoelectric bending in [110] oriented $DyScO_3$ inside an electron microscope. We corroborate these observations with independent ex-situ measurements of the flexoelectric coefficient with a three-point bending setup. The relevant flexocoupling voltage was measured to be -42(2) V, which is higher than expected based upon current flexoelectric models.

*Corresponding Author, Email: L-marks@northwestern.edu

First observed in solids by Bursian et al. [1], the term flexoelectricity was coined in the field of liquid crystals [2] and subsequently adopted for solids [3] and also biological membranes [4]. The flexoelectric effect (the presence of a polarization due to a strain gradient) occurs in many materials [4-9]; these effects are more pronounced as the thickness of a material decreases.

Examples of this pervasive phenomena include flexoelectricity-driven imprinting [10-12], internal bias in thin films [13, 14], nanoferroics [15] and dead layers in ferroelectric thin films [16, 17]. Flexoelectric coupling can change domain walls and interfaces in ferroelectrics [18-29] and ferroelastics [30-32], control defects [33] and change nanoindentation hardness of ferroelectrics [34-36]. It can also impact dielectric properties [13], photocurrents [37] and phonon spectra [38]. A number of other potential applications including flexoelectric energy harvesting [39, 40], photonic crystals [41] and strain sensors [42, 43] have been reported.

Recent progress has allowed for the experimental measurement of flexoelectric coefficients via small mechanical changes [66] and piezoresponse force microscopy [18, 44-46]. The theoretical understanding of the effect has advanced at the quantum level due to dynamic polarization theory [47-50] and density functional theory (DFT) [51-53], as well as at the continuum level with elastic theory [9, 54-56]. Additionally, there have been a number of papers analyzing the mechanics at the micron [20, 57] and nano scale [8, 58-61] as well as in thin films [33, 62, 63]. Consequently, flexoelectricity has been studied in a number of oxide systems [64-67].

In their pioneering work, Bursian et al. [1] observed curvature in a $BaTiO_3$ film due to the application of an electric field. A 2.5 μm thick film of $BaTiO_3$ was measured to have a curvature of 150 $m^{-1}$; it was predicted that in films of nanometer thickness, the curvature would be of the order of $10^6 - 10^{10}$ $m^{-1}$ [1]. While none of the existing literature includes direct experimental evidence demonstrating such high curvature values, there is one indirect measurement of comparable curvatures [19].

In this paper, we report direct experimental observation of large flexoelectric bending at the nanoscale in dysprosium scandate ($DyScO_3$) with similar results for two other lanthanide scandates ($TbScO_3$ and $GdScO_3$). Within a transmission electron microscope (TEM), thin rods of these single crystal oxides bend up to 90° with a radius of curvature ~10 μm when charged positive by the loss of secondary electrons. The bending is proportional to electron beam flux and not associated with any dislocations, twinning or similar changes in the oxide. We argue that the bending is due to a combination of a number of different factors: a well ordered valence compensated surface, a low density of states below the vacuum level, and a large flexocoupling voltage.

TEM samples were made from commercially-purchased [110] oriented single crystals of $DyScO_3$ using conventional mechanical polishing and ion beam thinning. The thinned samples were annealed in air for 12 hours at 1050 – 1200°C (see Methods Section for more details). TEM characterization was complimented with atomic force microscopy (AFM), X-ray photoelectron spectroscopy (XPS), ultraviolet photoelectron spectroscopy (UPS), reflection electron energy loss spectroscopy (REELS), and the surface structure was theoretical modelled using exact-exchange functionals (Methods and Supplemental Material).

As illustrated in Figure 1, when the electron beam was converged onto a thin rod of $DyScO_3$, it charged positively and bent downwards; if the beam was not centered on the rod it bent down and to one side.

Figure 2 shows a number of frames from a video (Supplemental Video V1) where the electron beam was defocused from (b) to (h) on a thin $DyScO_3$ rod. The rod charged and bent rapidly and returned to its original unbent form as the electron flux was reduced. The angles as well as the electron fluxes are given in Figure 2. The bright Fresnel fringes at the edge of the sample indicate downward bending. The electron fluxes were 1-100 electrons/nm$^2$s, significantly lower than $10^4$-$10^6$ electrons/nm$^2$s typically used for high resolution imaging. In many cases, the process was reversible although with too severe bending, the rods could fracture (Supplemental Video V2). The phenomenon was observed for about 50 different rods from 20 different samples, including $TbScO_3$ and $GdScO_3$ samples. We did not observe any dependence upon the crystallographic direction of the rods with the caveat that the thin direction was always [110]. The process appeared to be elastic with no evidence of dislocations or phase transitions. The bending was often on the time scale of the video recordings, suggesting adjustments of the charge took 10-1000 msec, although this does not preclude the possibility of faster processes. We looked for similar charging and bending in $SrTiO_3$, $KTaO_3$, $NdGaO_3$ and $LaAlO_3$. There is some charging in $NdGaO_3$ and $LaAlO_3$ (less than the scandates by more than an order of magnitude), little to no bending.

A second set of results shown in Figure 3, also frames from a video, depict the effects of asymmetric sample illumination (Supplemental Figure S1 and Supplemental Video V3). In this case, the rod reversibly bent away from the beam. For high incident electron energies charging will be net positive, involving the loss of secondary electrons. As discussed further in the Supplemental Material, the results in Figures 2 and 3 are consistent with more positive charge on the top surface of the sample closest to the beam center than on the bottom surface, resulting in a non-zero electric polarization between the two surfaces.

To verify that the observed bending was not associated with an electron optics artifact, we collected diffraction patterns while converging the beam as shown in Figure 4 (Supplemental Video V4). After the beam was focused, the sample was tilted by 24° to the [110] zone axis. The sequence in Figure 4 (a) to (h) depicts defocusing the beam (i.e. lowering the electron flux), and shows tilting of the sample by a total of 15.8°. While there was a slight deflection of the beam, it was several orders of magnitude smaller than the bending of the sample. For completeness, changing the microscope focus did not lead to large shifts in the beam, which is consistent with minimal bending of the electron beam.

A second possibility is that the ~2 T magnetic fields in the microscope play a role in the observed bending. $DyScO_3$ transitions from a paramagnetic to an antiferromagnetic state below 4 K and measurements have shown that a magnetic field of a few T can rotate $DyScO_3$ samples around this transition temperature [68]. Therefore, we performed experiments in electron microscopes that exposed the samples to ~5 Oe magnetic fields and observed similar bending (Supplemental Figure S2), ruling out magnetic contributions as the dominant term.

A third possibility is that the phenomenon depends upon the presence of occupied minority 4f states at the valence band maximum. Whereas $DyScO_3$ has two minority 4f electrons, $TbScO_3$ has one and $GdScO_3$ has none. Since all three showed approximately the same bending, the minority 4f states may play a role, but their presence is not required. A number of other remote possibilities such as Coulomb repulsion between charges on the surfaces are briefly discussed in the Supplemental Material. We find that they are orders of magnitude too small to account for the observed bending.

The diffraction patterns (Figure 4) show a low diffuse background with no evidence of additional reflections indicating few bulk defects and a well-ordered 1×1 surface. Based on the annealing conditions, we expect this surface to be valence neutral. AFM imaging (Supplemental Figure 3) confirmed that the surface was flat with monatomic steps of height 0.15 nm. The 1×1 [110] surface of $DyScO_3$ is similar to the 2×2 [001] surface of a simple perovskite. Angle resolved XPS measurements indicated that the surface was Sc rich with two Sc atoms per 1×1 surface cell (Supplemental Figure 3), similar to the well-established double-layer reconstructions on $SrTiO_3$ [001]. DFT calculations for different surface configurations indicated that the lowest energy structure contained three rows of scandium oxide along [001] as shown in Supplemental Figure S4. Additionally, XPS measurements were used to determine if water was present on the $DyScO_3$ TEM samples. Results indicated that the annealed sample prior to TEM experiments had minimal chemisorbed hydroxide (Supplemental Figure S5).

To understand why these samples charge more than any other we are aware of at very low electron fluxes, we utilized a combination of UPS and REELS to examine the electronic structure. UPS results (Supplemental Figure S6) indicated that the material had a bulk work function of 5.8 eV. The band gap, experimentally measured with REELS (Supplemental Figure S6), was found to be $5.7 \pm 0.1$ eV. The DFT calculations indicate that except for a small density of unoccupied states associated primarily with one of the surface Sc atoms and some upwards band bending near the surface, there was nothing unusual in the band structure. This is consistent with the experimental UPS and REELS data. These results explain the severe charging: secondary electrons produced by inelastic scattering in the bulk have sufficient energy to escape and minimal bulk traps (except those where electrons have already been lost) to fall into.

Is charging necessary for the bending, or just something that also occurs? We coated samples with a thin layer of carbon, and also did a low energy ion-beam milling of them. In neither case was the charging as severe (it was non-existent with the carbon coating) and the bending was minimal to none. Therefore, the charging and bending in our experiments are linked.

Structural and electronic characterizations were all consistent with the sample bending as a result of polarization induced on the sample, i.e. a flexoelectric response. To fully establish the origins of the bending, ex-situ characterization of the bulk flexoelectric effect in $DyScO_3$ was carried out using a three-point bending method. A dynamic mechanical analyzer was used to induce an oscillatory strain gradient and the resulting polarization was measured as a current using a lock-in amplifier (Supplemental Material) [69]. Polarization versus strain gradient for a [110] oriented $DyScO_3$ sample is shown in Figure 5. To validate the setup, the flexoelectric coefficients of [100] and [110] oriented $SrTiO_3$ were measured to have magnitudes of 12.4 nC/m and 8.3 nC/m, respectively, consistent with literature values which range from 1 nC/m to 10 nC/m [69]. All measurements were found to be independent of static force and oscillatory frequency. The magnitude of the flexoelectric coefficient for [110] oriented $DyScO_3$ was measured to be $8.4 \pm 0.4$ nC/m, using the standard error for thirteen different measurements. The signs for [110] oriented $DyScO_3$ and $SrTiO_3$ were consistently negative. Although the flexoelectric coefficient in $DyScO_3$ is of the same order of magnitude as the measured flexoelectric coefficient in $SrTiO_3$, the dielectric constant of $DyScO_3$ is an order of magnitude smaller than that of $SrTiO_3$ [70]. Therefore, the flexocoupling voltage (ratio of the flexoelectric coefficient to the dielectric constant) is an order of magnitude larger in $DyScO_3$ than in $SrTiO_3$ at $42 \pm 2$ V, large compared to typical values (1-10 V range [7]). The direct measurements support our model that the flexoelectric effect is a significant contribution to the bending observed in the electron microscope.

To connect quantitatively to the experimental observations, as discussed in more detail in the Supplemental Material, with an isotropic elasticity approximation for the curvature [71] would require a surface charge density of $3\times10^{-2}$ electrons/nm$^2$. While this is about one order of magnitude larger than other existing measurements of charging of samples, the lanthanide scandates charged much more than any other oxide sample we are aware of and the isotropic elasticity approximation will not be very accurate.

Charging and bending of samples are frequently observed inside transmission electron microscopes, treated as useless artifacts and ignored. We suspect the existence of more interesting science in these processes; flexoelectric bending may be the rule within electron microscopes rather than the exception. The detailed physical origin of the large flexocoupling voltage is an open question, particularly issues such as the role of octahedral rotations, the strong and anomalous polar phonons in these scandates [72], any surface contributions as well as other less likely contributors such as non-collinear magnetism. We do not want to speculate further here, and leave this to the future.

**Acknowledgments**

The authors are indebted to Yimei Zhu of Brookhaven National Labs and Amanda Petford-Long of Argonne National Labs for heroic assistance with extremely hard to handle samples in their low magnetic field transmission electron microscopes. We would also like to thank Oleg Rubel for information on the Berry Phase calculations as well as unreleased versions of the code BerryPI. We thank Fabien Trans and Peter Blaha for discussions on the use of hybrid functionals in the WIEN2k code, as well as James Rondinelli and Kenneth Poeppelmeier for their scientific input on the materials. Ex-situ flexoelectric measurements were made possible using equipment from L. Catherine Brinson and Lincoln J. Lauhon. PK and CM acknowledge funding by the Department of Energy on Grant Number DE-FG02-01ER45945.

**Figure Legends**

**Figure 1**. **Illustration of sample's electromechanical response under the electron beam**. (**a**) The sample is not bending due to low beam current. (**b**) The sample is bending down with a focused beam centered on the sample and (**c**) the sample is bending down and sideways with the focused beam centered on one side of the sample.

**Figure 2. Downward bending of a [110] oriented DyScO$_3$ sample under the electron beam**. Eight frames taken from Supplementary Video V1 showing the decrease in the downward bending of the sample with a gradual spreading of the electron beam from (**b**) to (**h**). The approximate electron flux (electrons/nm$^2$s) calculated using a quantum yield of 0.2 is given in the bottom right corner of each frame. The corresponding bending angles, given at the top right corner of each frame, are calculated by taking the apparent lengths of the rod in each frame to be the projection of the unbent rod in frame (**a**) at low beam current.

**Figure 3**. **Sideways bending of a [110] oriented DyScO$_3$ sample under the electron beam**. Eight frames of transmission electron microscopy images with the electron beam focused to one

side of the sample. The beam is centered towards the top half of the frame in (**a**) and gradually shifted towards the bottom of the frame in (**a**) through (**h**). The center of the beam is shown in Supplementary Figure S1, and the full data in Supplementary Video V3.

**Figure 4**. **Demonstration of bending of a [110] oriented sample of DyScO$_3$ via electron diffraction**. Eight frames of a transmission electron diffraction pattern (Supplementary Video V4) with the electron beam being defocused (reduced flux) from (**a**) – (**h**) showing the bending of the sample in reciprocal space. There is no discernable change (±0.1 nm$^{-1}$) in the distance between the transmitted beam and the mouse pointer which was used as a stationary reference point. A solid white arrow in (**e**) – (**h**) is drawn from the transmitted beam to the center of the approximate Laue circle. Tilt angles from the [110] zone axis in degrees are given in the bottom right corner of each frame based upon fitting circles to the strong spots on the Laue circle.

**Figure 5**. **Measured values of polarization as a function of strain gradient in [110] oriented DyScO$_3$.** Polarization versus strain gradient for a series of measurements performed at different oscillatory frequencies. The dotted line indicates a linear fit between strain gradient and polarization. The slope of such a line is the flexoelectric coefficient.

**Figure**

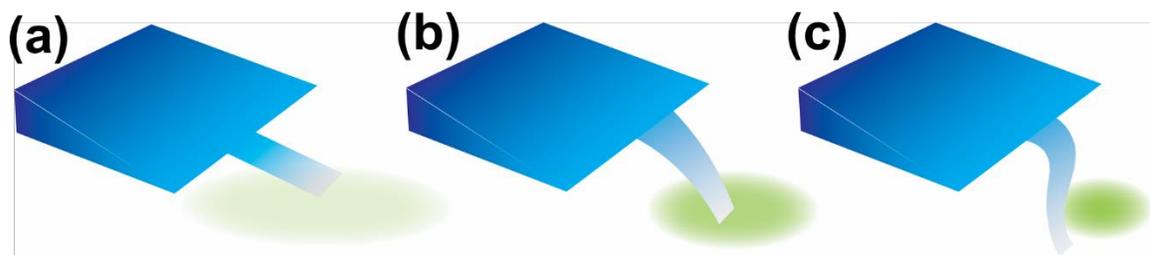

**Figure 1. Illustration of sample's electromechanical response under the electron beam.** (a) The sample is not bending due to low beam current. (b) The sample is bending down with a focused beam centered on the sample and (c) the sample is bending down and sideways with the focused beam centered on one side of the sample.

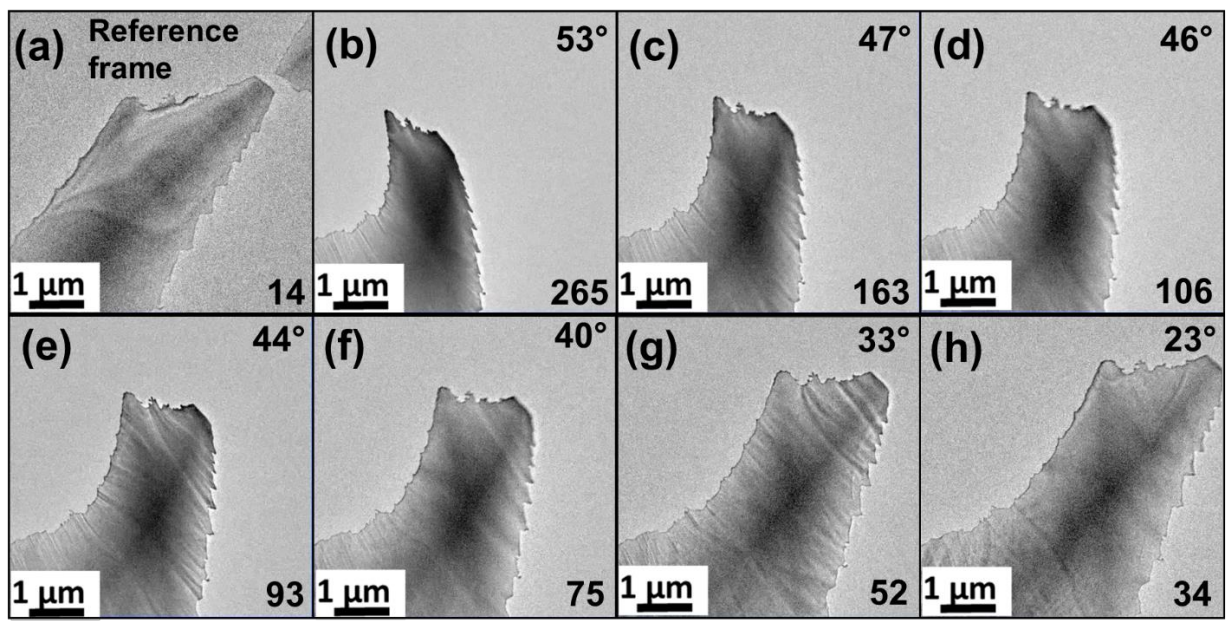

**Figure 2. Downward bending of a [110] oriented DyScO$_3$ sample under the electron beam.** Eight frames taken from Supplementary Video V1 showing the decrease in the downward bending of the sample with a gradual spreading of the electron beam from (b) to (h). The approximate electron flux (electrons/nm$^2$s) calculated using a quantum yield of 0.2 is given in the bottom right corner of each frame. The corresponding bending angles, given at the top right corner of each frame, are calculated by taking the apparent lengths of the rod in each frame to be the projection of the unbent rod in frame (a) at low beam current.

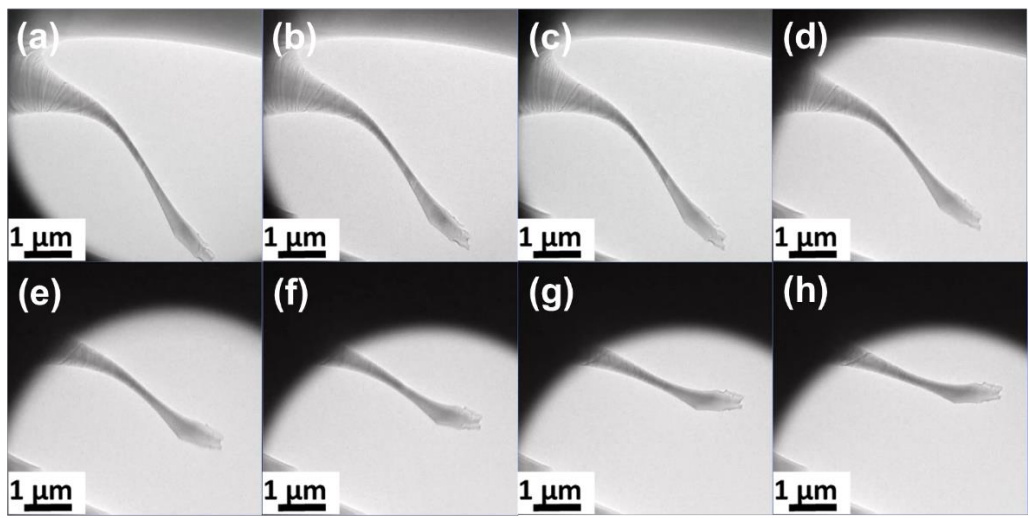

**Figure 3. Sideways bending of a [110] oriented DyScO₃ sample under the electron beam.** Eight frames of transmission electron microscopy images with the electron beam focused to one side of the sample. The beam is centered towards the top half of the frame in (**a**) and gradually shifted towards the bottom of the frame in (**a**) through (**h**). The center of the beam is shown in Supplementary Figure S1, and the full data in Supplementary Video V3.

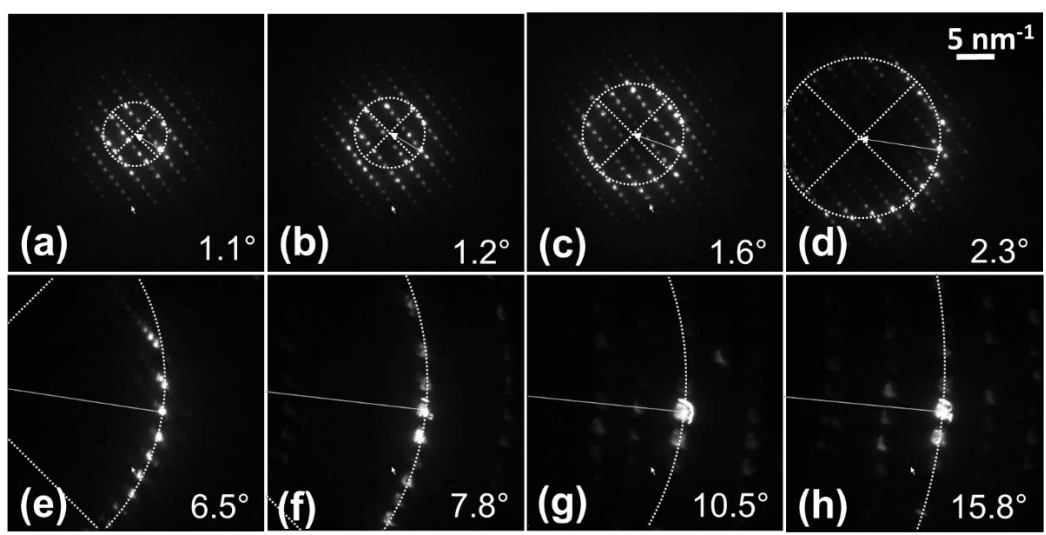

**Figure 4. Demonstration of bending of a [110] oriented sample of DyScO₃ via electron diffraction.** Eight frames of a transmission electron diffraction pattern (Supplementary Video V4) with the electron beam being defocused (reduced flux) from (**a**) – (**h**) showing the bending of the sample in reciprocal space. There is no discernable change (±0.1 nm⁻¹) in the distance between the transmitted beam and the mouse pointer which was used as a stationary reference point. A solid white arrow in (**e**) – (**h**) is drawn from the transmitted beam to the center of the approximate Laue circle. Tilt angles from the [110] zone axis in degrees are given in the bottom right corner of each frame based upon fitting circles to the strong spots on the Laue circle.

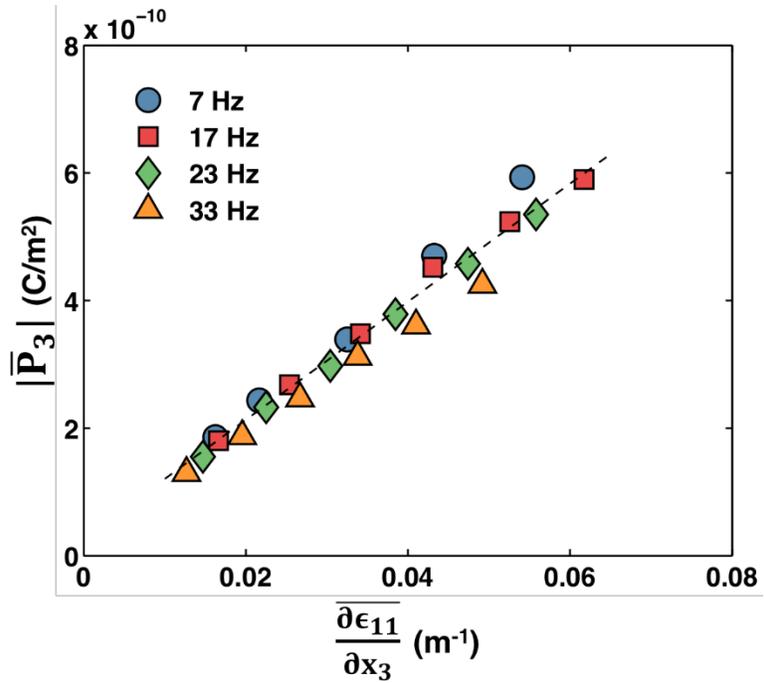

**Figure 5**. **Polarization versus strain gradient for a series of measurements performed at different oscillatory frequencies for [110] oriented DyScO₃**. The dotted line indicates a linear fit between strain gradient and polarization. The slope of such a line is the flexoelectric coefficient.

**Supplemental Material**

**Direct Observation of Large Flexoelectric Bending at the Nanoscale in Lanthanide Scandates**

Pratik Koirala, Christopher Mizzi and Laurence D. Marks

Department of Materials Science and Engineering, Northwestern University, Evanston, IL 60208, USA

1. **Materials and Methods:**

**TEM Sample Preparation**

Commercially available single crystalline substrates (MTI Corp, Richmond, CA) of $RScO_3$ (R=Dy, Tb, Gd) were cut into 3mm discs using an ultrasonic cutter, mechanically thinned to a thickness of ~100 µm using silicon carbide sandpaper, then dimpled with a Gatan 656 Dimple Grinder and 0.5 µm diamond slurry until the thickness at the center was ~15 µm. The samples were then $Ar^+$ ion milled to electron transparency using a Gatan Precision Ion Polishing System (PIPS-I) starting at an energy of 5 keV and milling angle of 10º. The ion energy and milling angles were gradually brought down to 3 keV and 4º respectively for final polishing and surface cleaning. Finally, the samples were annealed in a tube furnace for 8 – 12 hours at 1050 – 1200 ºC in air. Similar conditions are used to prepare samples with surface reconstruction on other oxides, particularly $SrTiO_3$ [1-9].

**X-ray Photoelectron Spectroscopy (XPS), Ultraviolet Photoelectron Spectroscopy (UPS) and Reflection Electron Energy Loss Spectroscopy (REELS)**

*Experimental Setup/Parameters*

XPS, UPS, and REELS measurements were taken on a multisource ESCALAB 250Xi. A monochromated, micro-focused Al K-Alpha x-ray source of 500 µm spot size and a 180° double focusing hemispherical analyzer with a dual detector system was used for XPS. An $Ar^+$ flood gun was used to compensate for charging and adventitious carbon was used to correct for any charging artifacts. A beam voltage of 2 V, emission current of 50 µA, focus voltage of 20 V, and extractor bias of 30 V was used on the flood gun. The Ar partial pressure in the chamber was ~$10^{-7}$ mbar (7.5 x $10^{-8}$ Torr). The hemispherical analyzer was located directly above the sample, and the sample stage rotated for angle resolved X-ray photoelectron spectroscopy measurements. The collection angles were measured with respect to the sample surface normal.

UPS spectra were taken using a high photon flux (>1.5 x $10^{12}$ photons/second) UV source with a spot size of approximately 1.5 mm. The source energy was 40.8 eV (He II).

REELS measurements were taken using a 1 keV incident energy. The emission current was stabilized at 5 µA and a 150 µm aperture was used. A pass energy of 10 eV was used with steps of 0.1 eV and 50 ms dwell time.

*XPS Fitting*

XPS fitting was done using the Powell method [10] to deconvolve the peaks using a fixed Gaussian-Lorentzian (GL) mixing parameter of 30%, which is critical for addressing the asymmetries in XPS spectra. A maximum of 20,000 iterations was used to converge the fit to $1\times10^{-6}$ counts per second.

The Dy3d doublets were fitted with the following constraints [11]:

1. Area of Dy $3d^{3/2}$ = 0.7 × Area of Dy $3d^{5/2}$
2. Peak binding energy of Dy $3d^{3/2}$ = Peak binding energy of Dy $3d^{5/2}$ + 38 (±0.1) eV

The Sc2p doublets were fitted with the following constraints [11, 12]:

1. Area of Sc $2p^{1/2}$ = 0.50 × Area of Sc $2p^{3/2}$
2. Peak binding energy of Sc $2p^{1/2}$ = Peak binding energy of Sc $2p^{3/2}$ + 4.5 (±0.1) eV

*XPS Model*

XPS data was fitted with a layer model [13] taking into account the relative photoionization cross-section, inelastic mean free path, and relative atom density (atoms/cm³) of the different species. The intensity of the Sc 2p peak taking into account the damping in the immediate $DyO^+$ layer is

$$I_{Sc} = F\alpha_{Sc} D_{Sc} k\lambda_{Sc} e^{\frac{-t}{g\lambda_{Sc}}} \quad (1)$$

where F is the flux of the incident radiation, $\alpha$ is the photoionization cross-section, k is a spectrometer factor, g = cos θ (where θ is the take-off angle measured from the surface normal), t is the spacing along [110], D is the atomic density in the (110) plane, and λ is the appropriate inelastic mean free path. Similarly, the intensity of Dy 3d taking into account the damping in the immediate $ScO_2$ layer is

$$I_{Dy} = F\alpha_{Dy} D_{Dy} k\lambda_{Dy} e^{\frac{-t}{g\lambda_{Dy}}} \quad (2)$$

When fitting a layer model to the experimental data, only the relative intensities are important, and can be written as:

$$\frac{I_{Sc}}{I_{Dy}} = \frac{\alpha_{Sc}\lambda_{Sc} e^{\frac{-t}{g\lambda_{Sc}}}}{\alpha_{Dy}\lambda_{Dy} e^{\frac{-t}{g\lambda_{Dy}}}} \quad (3)$$

The intensities were then integrated over a total thickness corresponding to three times the inelastic mean free paths of the corresponding elements. Relative intensities were calculated for different surface concentrations and normalized to the experimental data at 0° take off angle, measured with respect to the surface normal. The relative intensities rather than absolute numbers are the correct terms to compare because many factors such as chemisorbed species on the surface or carbon contamination incurred during sample transfer from the annealing furnace to the XPS chamber can change absolute intensities.

**Atomic Force Microscopy (AFM)**

Tapping mode AFM imaging was done using a Bruker's *Dimension FastScan* in air. 1024×1024 pixels were scanned for a 2×2 µm² area of the sample.

**Transmission Electron Microscopy (TEM)**

Transmission electron microscopy was performed using a Hitachi H8100 operated at 200 kV. A nominal exposure time of 0.1 seconds was used with an electron flux in the range of 1- 100 electrons/nm$^2$s on the sample. The bending experiments were done starting with the beam spread out and gradually converged, although for recording the videos it was often easier to start with a focused beam and defocus it.

To investigate the effect of a high magnetic field on the specimen, similar experiments were performed in a low field JEOL2100F at Argonne National Lab. Experiments similar to the one performed on a H8100 described above were carried out. The electron beam was gradually converged starting with a highly spread out beam. It was exceedingly difficult to stabilize the sample which pulsated periodically due to the field emission source. However, on converging the beam it stabilized, at which point it appeared that the charge saturated. As a result, a small window of beam convergence was found where it was possible to perform the bending experiments. With the field emission source it was not possible to do electron diffraction.

## 2. Bending Estimation from Transmission Electron Diffraction

A video was recorded taking transmission electron diffraction patterns with a gradually focused electron beam. Frames at different time intervals were used to estimate the bending.

An approximate Laue circle was drawn tracing the reflections with higher intensity compared to the surrounding reflections. Subsequently, a line drawn from the direct beam to the center of the approximate circle was used to estimate radius of the circle (r) (see Supplemental Figure S4) and hence the amount of bending. The bending angle, $\theta$, in radians is

$$\theta = \sin^{-1}(\lambda r) \qquad (4)$$

where $\lambda$ is the wavelength of the 200 keV electrons in nm and $r$ is the radius of the approximate Laue circle in nm$^{-1}$.

## 3. Density Functional Theory (DFT) Calculations

DFT calculations were performed with the all-electron augmented plane wave + local orbitals WIEN2K code [14]. Muffin tin radii of 1.68, 1.82, and 2.02 were used for O, Sc and Dy, respectively, to minimize inclusion of tails of the O 2p density perturbing the calculation of the exact-exchange corrections inside the muffin tins for Sc and Dy. The plane-wave expansion parameter RKMAX in the code was 6.5. Atom positions and bulk optimized lattice constants were calculated using the on-site hybrid method [15, 16] with the PBESol functional [17]. For the surface a 70.000x7.925x7.9357 Å cell was used containing 260 atoms (92 unique) with P121/m1 symmetry and a 4x4x1 mesh. The electron density and atomic positions were simultaneously converged using a quasi-Newton algorithm [18]; the numerical convergence was better than 0.01 eV (1×1 cell)$^{-1}$ surface cell. All calculations were with ferromagnetic unit cells, which is appropriate for the samples in an electron microscope; the difference in positions for ferromagnetic and anti-ferromagnetic ordering was minimal, as would be expected since this is a weak energy term.

Born charges calculated using the BerryPI package [19] in WIEN2k were consistent with the existing *ab initio* literature [20], showing nothing anomalous with effective charges slightly larger (10-30%) than the nominal valences of $Dy^{3+}$, $Sc^{3+}$ and $O^{2-}$.

A complicated issue with $DyScO_3$ as well as the other lanthanide scandates, is that there is relatively little experimental information on the band structure, and uncertainties in the lattice parameters and atomic positions. As discussed in the recent literature [21], the lattice parameter measured by different groups differs more than one expects, and coupled with slight variations in atomic positions corresponds to a variation of $0.1 - 0.2$ in the bond valence values which is high. We therefore used a slightly complicated strategy to ensure that our simulations provided an adequate representation. First, as is well known, simple DFT functionals give incorrect results for 4f electrons and can lead to over hybridization of the Sc 3d and Dy 5d states with the O 2p states. We note that while a correction for the Dy 5d states is not common, if this was omitted the Dy-O distances were much too small. We varied the fraction of on-site exchange for the Sc 3d and Dy 4f and 5d to obtain a result which gave positions close to those experimentally observed, i.e. minimized the forces.

On-site corrections of 0.80 for the Sc 3d and 0.5 and 0.30 for the Dy 5d and 4f were approximately optimum. The values of the corrections for the d electrons were critical to obtaining atomic positions close to those found experimentally; that for the 4f were not. Inclusion of spin-orbit coupling was tested and while this changed fine details of the electronic structure for the 4f electrons, it was otherwise insignificant.

The procedure for the surface calculations was to minimize the energy for all possible surface combinations of the Sc positions for the experimentally determined coverage and a fully oxidized surface using identical parameters with the on-site method. The structure in the Supplemental CIF file was significantly lower in energy than any others.

For completeness, we will mention that there is considerable ambiguity about the 4f minority electrons in $DyScO_3$ and $TbScO_3$, particularly exactly where they lie relative to the Fermi level. As mentioned above, the 4f correction has little effect upon the atomic positions so cannot be independently determined. We have extensive evidence from more detailed XPS analyses as a function of temperature including the presence of a surface insulator to metal transition above about 90C for $DyScO_3$ and 220 C for $TbScO_3$ which involves the 4f minority electrons, which we will report in more detail elsewhere. Since comparable bending was observed for $GdScO_3$ which does not contain any minority 4f, we conclude that they play no role in the flexoelectric contributions so this other science while interesting in its own right is not relevant here.

## 4. Additional Material on Charging in the Electron Microscope

High energy electrons as used in an electron microscope are a white source for inelastic scattering, leading to everything from phonon to plasmon, core and Bremsstrahlung processes. At the much lower accelerating voltages of a scanning electron microscope there can be a net deposition of electrons into the sample, but above a (material and sample specific) energy more secondaries are lost than electrons trapped from the beam. The samples will be net positive through loss of secondary electrons, although the charge distribution does not have to be simple.

There are two broad classes of secondary electrons:
  a) SE1, which escape directly from the sample following an inelastic scattering event.

b) SE2, which are generally delocalized and involve additional scattering of electrons after the initial scattering event, and the latter can take place some distance from the surface.

The SE1 electrons can have atomic resolution [8, 22], and calculations [23] indicate approximately equal probabilities from both the entrance and exit surface. The dominant process for SE2 is probably plasmon excitations, produced by the electrostatic shock wave of the high energy electron going through the sample. While the majority of the momentum transfer will be normal to the electron beam, there will in general be an average component along the beam direction. The sample was extremely unstable due to the large charging, so direct measurement of the amount of charging or the potential was not possible.

To compare the experimental bending with theoretical expectations, the curvature can be estimated [24] as:

$$\xi = f \frac{12\sigma(1 - \vartheta^2)}{Gd^2} \qquad (5)$$

where $\xi$ is the curvature, f is the flexocoupling voltage, G is the Young's modulus, $\sigma$ is the surface charge density, d is the film thickness and $\vartheta$ is the Poisson's ratio. We note that the scandates have quite anisotropic elastic constants [25], so this form is not exact.

A number of experiments [26-36] have been performed using electron holography including a recent study on MgO smoke particles [37]. MgO particles were found to have a surface charge density of about $1.751 \times 10^{-3}$ electrons/nm$^2$. Assuming an identical field as would be produced by the amount of surface charge density measured for MgO smoke particles along with the experimentally measured value of the flexocoupling voltage f=42 Volts [38], yields a curvature of about $1.4 \times 10^3$ m$^{-1}$, which is one order of magnitude smaller than the measured valued of $3 \times 10^4$ m$^{-1}$ for DyScO$_3$. Experimentally the amount of charging of DyScO$_3$ was substantially larger than for comparable well annealed single crystals MgO [39]; indeed, the well-annealed DyScO$_3$ samples charged more than any sample we have encountered and were capable of retaining charge even when removed from the microscope into air (Supplemental Video 5). Inverting equation (5) above the surface charge density on the sample is calculated to be 0.03 electrons/nm$^2$. We believe this is not an unreasonable estimate.

## 5. Sign of the Flexoelectric Coefficient and the Charging

Considering the phenomenological definition of the flexoelectric coefficient:

$$P = \mu \frac{de}{dz} \qquad (6)$$

where z is the direction of bending, $e$ is the strain in the sample, $\mu$ is the flexoelectric coefficient, and $P$ is the polarization. When we illuminated the sample, it bent both down and away from the beam. Based upon the previously given secondary electron argument, the illuminated surface should charge positive with respect to the bottom surface. Since polarization is defined as a vector pointing from a positive charge density to a negative charge density, $P$ must point in the opposite direction of the strain gradient. Hence, the sign of the flexoelectric coefficient is expected to be negative. This is consistent with the sign of the flexoelectric coefficient from the ex-situ measurements using the three-point bending setup.

## 6. Other Contributions

There are a number of other possible contributing terms which are too small to be dominant, although they might play a small role.

The first is the contribution of the electrostatic repulsion of charges on the surface of the sample to the observed bending. As an approximation of the surface charge density, the values from above will be used, i.e. a charge density of $3 \times 10^{-2}$ electrons/nm$^2$. The contribution of the surface charge density to the net surface energy term can be approximated as a two dimensional screened Coulomb sum including a conventional surface dielectric screening:

$$U = \frac{1}{4\pi\varepsilon_0} \frac{2}{(1+\varepsilon_r)} \sum_{i,j \neq 0, i,j \to -\infty}^{\infty} \frac{e^2}{a\sqrt{i^2+j^2}} \cdot \exp\left(-\frac{a\sqrt{i^2+j^2}}{\lambda}\right) \quad (8)$$

where, a is the separation of charges, $\lambda$ is the screening length, i and j are the set of integers.

Approximating $\lambda$ to be of the order of tens of nanometers, the electrostatic contribution to the energy is of the order of 3 mJ/m$^2$. The typical number for surface energy of oxides is of the order of few J/m$^2$, hence the contribution of the surface charge density to the observed bending is negligible. To put this into context, the strain in the material due to this term can be compared to the effect of surface stress on the change in the lattice parameters of a small particle. The surface stress is of the order of the surface free energy, and for a 30 nm diameter nanoparticle the typical change in lattice parameter is about $2\times10^{-3}$ (see [40] and references therein). So the contribution of surface charging will be a strain of about $1\times10^{-6}$. For reference, we note that the Coulomb contribution can also be estimated in reciprocal space with the mean term omitted, which gives comparable results.

A second contribution is the volumetric change in lattice parameter due to a change in the local electron density at the surface associated with the charging. Based upon DFT calculations this corresponds to a fractional volume change of $8\times10^{-3}$ for a hole density of 1 hole per nm$^3$. With the above estimate of the charge density at the surface, this is negligibly small.

In addition to these, there are some terms which could matter although their magnitude and even existence is controversial. Surface contributions to the flexoelectric effect is a very controversial topic. There has been one specific calculation of the surface contributions to the flexoelectric effect to date [41]. There may also be contributions due to surface piezoelectric effects since the surface structure does not contain in-plane inversion symmetry. We will leave this to future research. We will also leave to the future calculations of the flexoelectric effect from first principles which is not an easy task and to date has often yielded results substantially smaller than experiment [42, 43].

## 7. Ex-situ Flexoelectric Measurements

Flexoelectric coefficient measurements were performed using a three-point bending configuration similar to the one described in Zubko et al [44]. Commercially available single crystalline substrates (MTI Corp, Richmond, CA) were cut into flexoelectric samples with typical dimensions of 10x10x0.5 mm and 10x5x0.5 mm. Approximately 50 nm Au electrodes were deposited using a sputter coater. Ag paste was used to attach Cu wires to the Au electrodes. Samples were annealed at 300 °C for approximately 2 hours to improve mechanical stability and electrical conductivity. Measurements were made with and without scraping wire to expose Cu from oxidized Cu. This had minimal impact on measurements.

A dynamic mechanical analyzer (DMA) was used to bend the sample at a specified frequency. A variety of static forces and frequencies were used (typically in the 10-40 Hz range). Static forces were chosen to be small enough to avoid any piezoelectric contributions. The sample sat on two alumina rods spaced 8.4125 mm apart which were held in a custom machined sample holder. A lock-in amplifier was used to measure the current that was generated due to the flexoelectric effect. Displacements were calculated using the elastic moduli [25, 45] and force measurements from the DMA. Flexoelectric coefficient as a function of different static forces and oscillation frequencies for [110] oriented $DyScO_3$ given in Supplemental Table T1. This clearly shows that the flexoelectric coefficient is independent of both of static force and oscillation frequency in our measurement regime.

**Supplemental Figures**

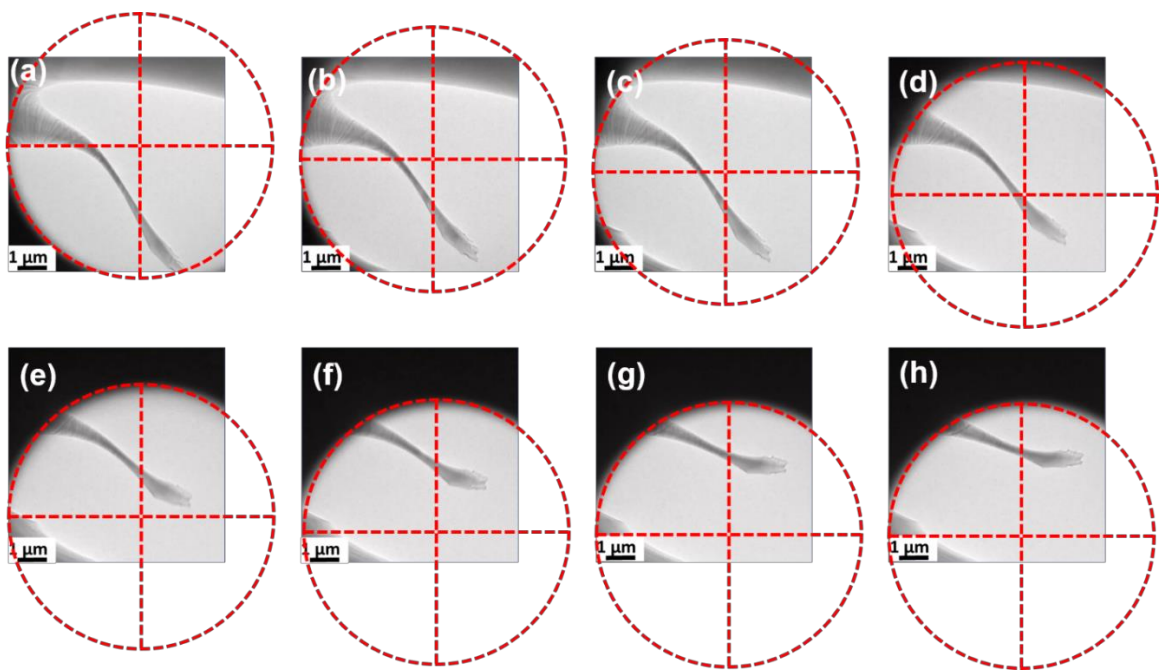

**Supplemental Figure S1.** Eight frames from the video with the electron beam focused more on one side of the sample. The beam is centered towards the top half of the frame in (**a**) and gradually shifted towards the bottom of the frame in (**a**) through (**h**). The red dotted circle is marked in each frame to indicate the approximate center of the electron beam illumination.

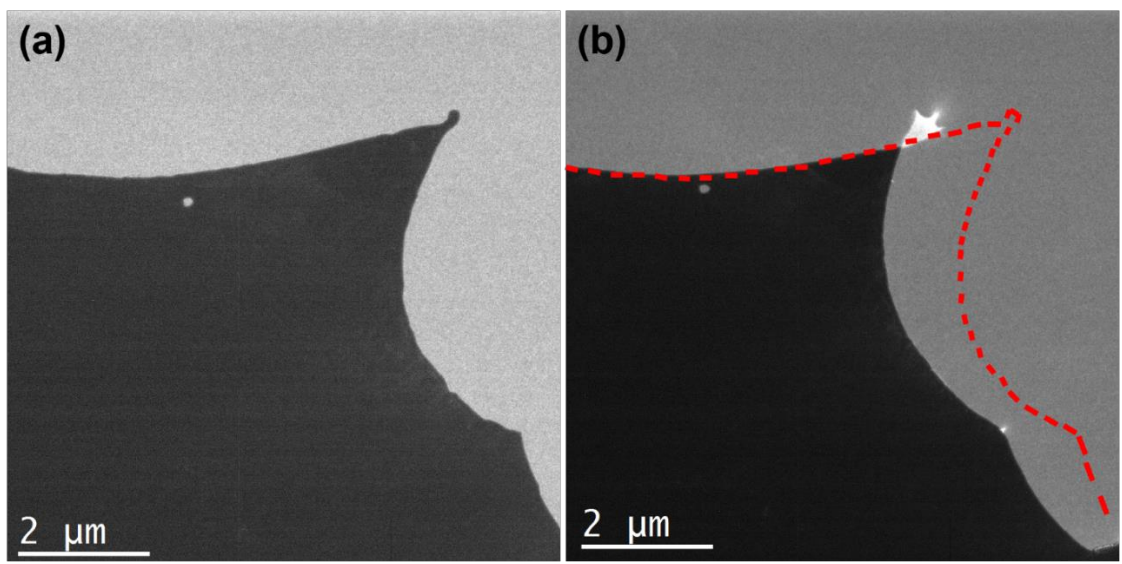

**Supplemental Figure S2.** Transmission electron microscopy images of DyScO₃ performed in a low field (~ 5 Oe) microscope showing the impact of beam convergence from (a) to (b). Dashed red line in (b) traces the edge of the sample prior to beam convergence.

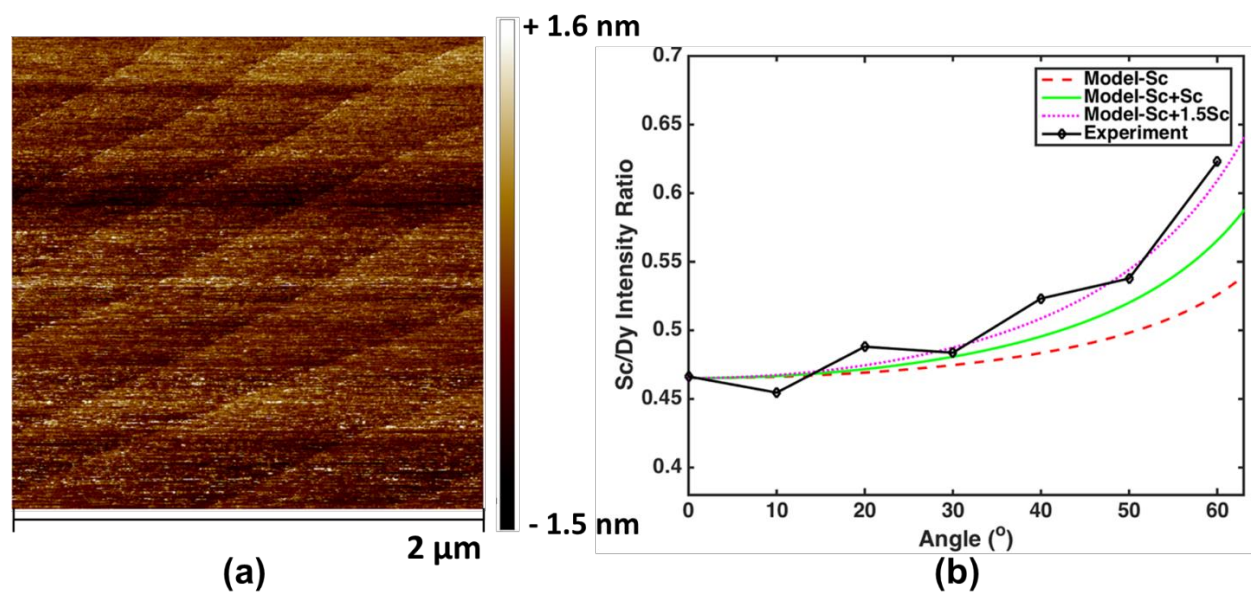

**Supplemental Figure S3.** Atomic force microscopy (AFM) image (in tapping mode) of a 2μm × 2μm area of a self-supporting (3 mm diameter) transmission electron microscopy (TEM) sample showing flat steps and terraces in **(a)**, and angle resolved X-ray photoelectron spectroscopy experimental data and fit in **(b)**.

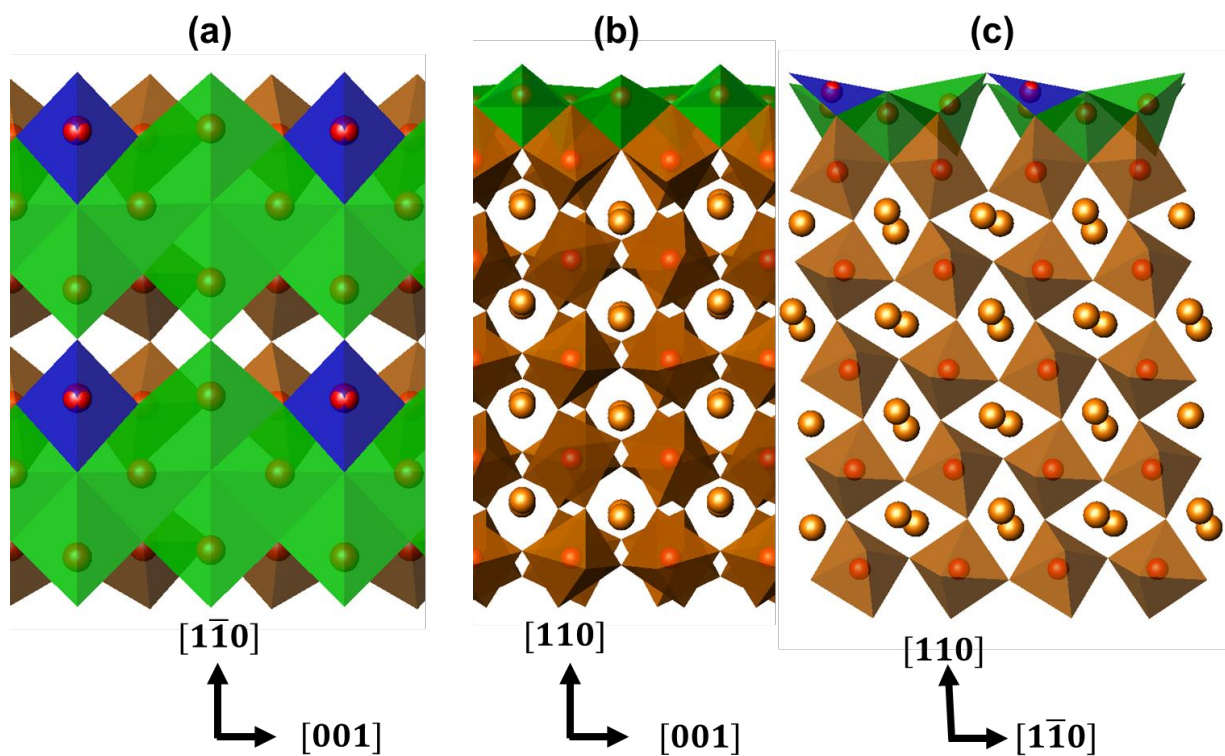

**Supplemental Figure S4.** Top of the DFT relaxed structure of DyScO$_3$ with 2.5 surface ScO$_2$ layer from three different orientations. ScO$_4$ tetrahedra are in blue, ScO$_5$ octahedra with an unoccupied oxygen site are in green, and ScO$_6$ octahedra are in brown.

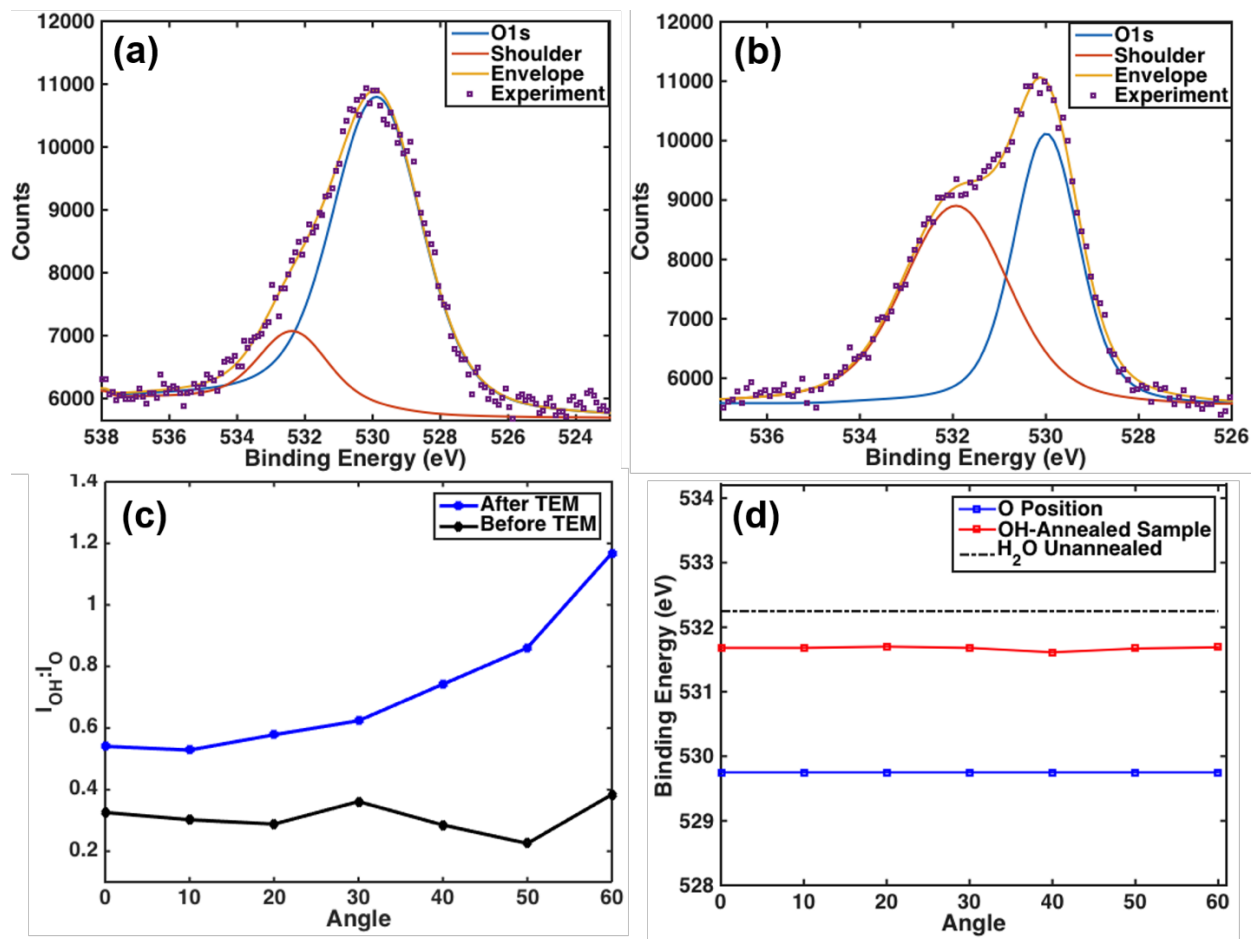

**Supplemental Figure S5.** In (**a**), XPS spectra of the O1s peak before TEM and (**b**) after TEM, both collected at 60º. Angle resolved XPS of the O1s shoulder to main peak intensity before and after TEM in (**c**) and the corresponding peak positions in (**d**). The dotted line in (**d**) marks the position of the shoulder for a sample with molecularly adsorbed $H_2O$ measured only at 0º tilt. The angles are measured with respect to the surface normal.

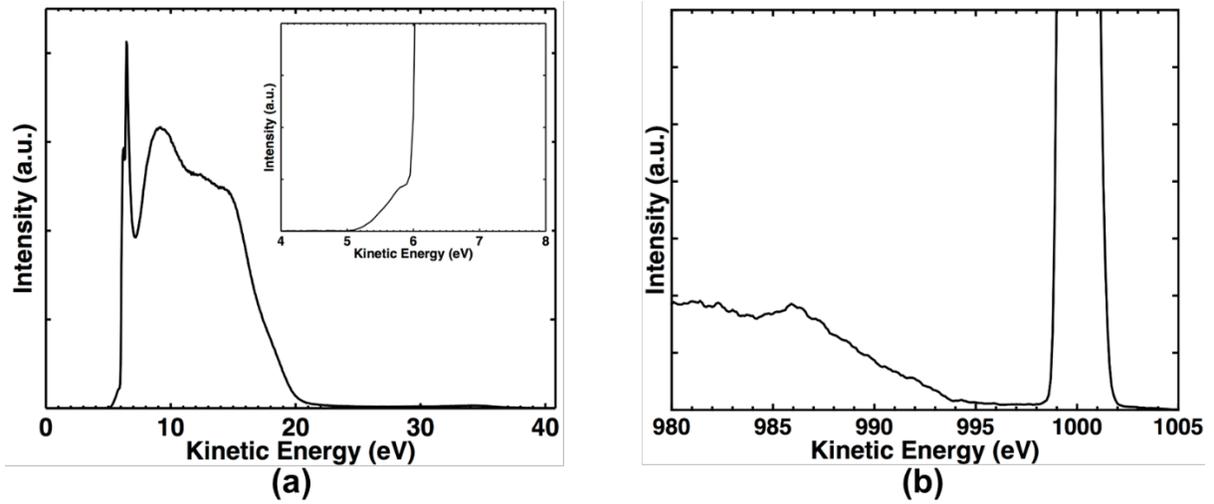

**Supplemental Figure S6.** (**a**) UV photoelectron spectrum (UPS) along with an embedded subpanel showing the fine details of the edge. (**b**) Reflection electron energy loss spectroscopy (REELS) with an incident energy of 1 keV showing no energy loss events up to the vacuum level.

**Supplemental Table T1**: Different measurements of the flexoelectric coefficient as a function of static force and oscillation frequency for [110] oriented $DyScO_3$.

| Flexoelectric Coefficient [nC/m] | Static Force [g] | Frequency [Hz] |
|---|---|---|
| 10.7 | 300 | 33 |
| 6.09 | 200 | 33 |
| 8.73 | 250 | 33 |
| 8.15 | 150 | 33 |
| 7.98 | 200 | 33 |
| 8.68 | 200 | 33 |
| 9.21 | 200 | 23 |
| 9.23 | 200 | 17 |
| 8.11 | 200 | 33 |
| 10.7 | 200 | 7 |
| 8.03 | 100 – 250 | 23, 33 |
| 5.57 | 100 – 200 | 33 |
| 7.49 | 150 | 33 |

**Supplemental Videos**

**Supplemental Video V1.** Real time video using conventional bright field imaging with the focus of the condenser lens changed to alter the electron flux on the sample. During the course of the video, the beam is first focused then defocused. Bend contours appear as lines across the rod due to the change in local orientation; there is no evidence for dislocations.

http://www.numis.northwestern.edu/Research/Projects/flexo/Supplemental%20Movie%20M1.mov

**Supplemental Video V2.** Real time video using conventional bright field imaging of a sample that bent by about ninety degrees and then fractured.

http://www.numis.northwestern.edu/Research/Projects/flexo/Supplemental%20Movie%20M2.mov

**Supplemental Video V3.** Real time video using conventional bright field imaging with a constant focus of the illumination. As the beam was moved from side to side, the thin rod bends away from the beam (it is already bent down).

http://www.numis.northwestern.edu/Research/Projects/flexo/Supplemental%20Movie%20M3.mov

**Supplemental Video V4.** Real time video using diffraction with the illumination initially focused after the sample had been tilted ~24º to the [110] zone axis. The cursor recorded in the video was not moved during the experiment.

http://www.numis.northwestern.edu/Research/Projects/flexo/Supplemental%20Movie%20M4.mov

**Supplemental Video V5.** Video after the sample had been removed from the microscope. When pushed with a tweezer it partially moves out, but when removed the sample returns to the cup. This suggests that charge has been stored in the sample during analysis, which is consistent with the much higher hydroxide chemisorption shown in Supplemental Figure S4.

http://www.numis.northwestern.edu/Research/Projects/flexo/Supplemental%20Movie%20M5.mov

**Supplemental CIF DyScO3_110.cif.** Conventional Crystallography Information File (CIF) of the surface slab used for the calculations.